\documentclass[conference]{IEEEtran}
\ifCLASSINFOpdf
\else
\fi
\usepackage{graphicx}

\usepackage{xfrac}
\usepackage{amsmath}

\usepackage{subfigure}
\usepackage{subfig}   

\usepackage{url}

\usepackage{multirow}
\usepackage{array}
\usepackage{tabularx}

\newcolumntype{P}[1]{>{\centering\arraybackslash}p{#1}}
\newcolumntype{M}[1]{>{\centering\arraybackslash}m{#1}}

\begin{document}
%
\title{Identifying highly influential travellers for spreading disease on a public transport system}

\author{
    \IEEEauthorblockN{Ahmad El Shoghri\IEEEauthorrefmark{1}\IEEEauthorrefmark{2}, Jessica Liebig\IEEEauthorrefmark{2}, Raja Jurdak\IEEEauthorrefmark{2}\IEEEauthorrefmark{3}, Lauren Gardner\IEEEauthorrefmark{4}\IEEEauthorrefmark{5}, Salil S. Kanhere\IEEEauthorrefmark{1}}
    \IEEEauthorblockA{\IEEEauthorrefmark{1}School of Computer Science and Engineering, University of New South Wales, Sydney, NSW, AUSTRALIA
    \\ahmad.elshoghri@student.unsw.edu.au}
    \IEEEauthorblockA{\IEEEauthorrefmark{2}Data61, Commonwealth Scientific and Industrial Research Organization, Brisbane, QLD, AUSTRALIA
    }
    \IEEEauthorblockA{\IEEEauthorrefmark{3}School of Computer Science, Queensland University of Technology, Brisbane, QLD, AUSTRALIA
    }
    \IEEEauthorblockA{\IEEEauthorrefmark{4}Department of Civil and Systems Engineering, Johns Hopkins University, Baltimore, MD, USA
    }
    \IEEEauthorblockA{\IEEEauthorrefmark{5} Research Center for Integrated Transport Innovation (rCITI), UNSW Sydney, Sydney, NSW, AUSTRALIA
    }
}


%



\maketitle
\IEEEpubidadjcol
\begin{abstract}
The recent outbreak of a novel coronavirus and its rapid spread underlines the importance of understanding human mobility. Enclosed spaces, such as public transport vehicles (e.g. buses and trains), offer a suitable environment for infections to spread widely and quickly. Investigating the movement patterns and the physical encounters of individuals on public transit systems is thus critical to understand the drivers of infectious disease outbreaks. For instance previous work has explored the impact of recurring patterns inherent in human mobility on disease spread, but has not considered other dimensions such as the distance travelled or the number of encounters.
Here, we consider multiple mobility dimensions simultaneously to uncover critical information for the design of effective intervention strategies. We use one month of citywide smart card travel data collected in Sydney, Australia to classify bus passengers along three dimensions, namely the degree of exploration, the distance travelled and the number of encounters. Additionally, we simulate disease spread on the transport network and trace the infection paths. We investigate in detail the transmissions between the classified groups while varying the infection probability and the suspension time of pathogens. Our results show that characterizing individuals along multiple dimensions simultaneously uncovers a complex infection interplay between the different groups of passengers, that would remain hidden when considering only a single dimension. We also identify groups that are more influential than others given specific disease characteristics, which can guide containment and vaccination efforts.
\end{abstract}


%
\IEEEpeerreviewmaketitle

\section{Introduction}

Human mobility continues to play a vital role in spreading infectious diseases within a population \cite{shahzamal2018impact}\cite{prothero1977disease}. Ongoing population growth and the high reliance of individuals on public transport services in highly populated cities provide a suitable platform for contagious diseases, such as measles, the recently emerged coronavirus and influenza, to spread widely and rapidly \cite{liu2019investigating}\cite{browne2016roles}\cite{publicTranspShutDown}. For example, individuals travelling on a bus are in close enough proximity to infect each other and can carry the infection to distant locations across the public transport network \cite{Nasir2016}\cite{browne2016roles}. Additionally, some pathogens may remain in the environment (e.g. a bus) for a prolonged period and can infect susceptible individuals after the infectious person has left the area \cite{shahzamal2018impact}\cite{shahzamal2019indirect}. Furthermore, transport services shorten distances and times and strongly connect different suburbs, potentially exposing communities to a high infection risk \cite{browne2016roles}. The risk of disease spread due to human movement is evident from the current novel coronavirus outbreak in China and internationally, with Chinese authorities shutting down public transportation within the affected area \cite{publicTranspShutDown}.

The recent uptake of smart travel cards and the availability of this data have created an unprecedented proxy to elicit different travelling behaviours and to study their effects on disease spread \cite{wesolowski2016connecting}\cite{bansal2016big}. The analysis of such data is critical to understand the spreading dynamics of a disease and consequently to develop effective containment strategies \cite{rubrichi2017comparison}. Previous studies investigated several spreading dynamics of infectious diseases, however, to the best of our knowledge none of the studies has incorporated different aspects and dimensions of mobility behaviour simultaneously. 

In this paper we study three aspects of mobility behaviour, i.e. the degree of exploration, the distance travelled and the number of encounters of passengers using the Sydney bus network in the context of infectious disease spread. By considering the three dimensions simultaneously, we identify previously unknown mobility behaviours. The high spatiotemporal resolution of the dataset allows us to construct a time resolved physical human contact network to simulate disease spread. Specifically, we trace the infection flows between groups of passengers who display different mobility behaviours to investigate the change in the spreading dynamics. In addition, we investigate how changes in the infection probability and the time pathogens remain suspended in the environment affect the spreading of the disease. This study identifies the most influential passenger groups in a disease spread scenario for different disease characteristics and types. Our simulation results identify four dominant transmission paths between the mobility groups that should become a focus of containment efforts. In addition, we find that highly connected passengers who regularly visit the same places have the highest spreading power when pathogens do not remain in the environment. However, with an increase in the suspension time of pathogens, highly connected passengers who visit new locations become the most efficient spreaders. An increase in the infection probability on the other hand, amplifies the spreading power of all mobility groups, especially for passengers who regularly visit the same places and travel short distances, until reaching a saturation point at a probability of 0.5.

The remainder of the paper is organised as follows: We begin by discussing relevant previous work in Section \ref{related work section}. In Section \ref{materials and methods} we present our framework for modelling infectious diseases on human contact networks. We explain the approach for classifying the individuals based on their movement behaviour and introduce the dataset that we use for our case study. In Section \ref{results and discussion} we run extensive trace driven simulations to investigate the underlying interactions and disease transmission dynamics between the different groups of passengers. Furthermore, we study the effect of changing the infection probability and the time pathogens remain in the environment on the transmission dynamics. Finally, we identify the most influential mobility behaviours for various disease characteristics, which can guide intervention strategies. Section \ref{conclusion and future work section} concludes the paper.

 

\section{Related Work} \label{related work section}
Studies of epidemiology have long recognized that human mobility plays a key role in fostering severe disease epidemics that may result in high rates of morbidity and mortality \cite{browne2016roles}. Furthermore, these studies have acknowledged the importance of identifying the most influential individuals, as it can aid to predict outbreaks before their occurrence \cite{rubrichi2017comparison}\cite{wesolowski2014commentary}. Health related datasets and detailed patient mobility profiles present informative data that may be used to reflect the status of a disease and its progression \cite{wesolowski2016connecting}\cite{nie2015beyond}. However, accessing such information is challenging due to privacy concerns and other related issues \cite{rubrichi2017comparison}\cite{nie2015beyond}.

In the absence of health related data, previous work has studied alternative data sources. An important body of research has explored the use of call detail records (CDRs) and data from the global positioning system (GPS) to build epidemiological models and to study the spatial transmission of various diseases in a population at both city and country levels \cite{wesolowski2012quantifying}\cite{tatem2014integrating}\cite{wesolowski2015impact}\cite{bengtsson2015using}\cite{wesolowski2015quantifying}\cite{isdory2015impact}\cite{brdar2016unveiling}\cite{pappalardo2015returners}.

In \cite{pappalardo2015returners} CDR and GPS datasets were exploited to extract two types of mobility behaviours. The authors used the recurrent mobility and the total mobility characteristics to group individuals into returners and explorers. Returners are individuals who can be characterized by their most visited locations as these dominate their movement behaviour, whereas explorers are individuals who often visit new places and cannot be characterized by their most frequently visited locations. The statistical measure used to compute the total mobility of an individual is the total radius of gyration $r_g$, defined as \cite{pappalardo2015returners}:
\begin{equation}
\label{eqn:total_radius_of_gyration}
r_g = \sqrt{ \frac{1}{N} \sum_{i \in L}{ n_i (r_i - r_{cm})^2 } },
\end{equation}
where $L$ is the set of all visited locations by the individual, $r_i$ is the coordinates of the visited location $i$, $n_i$ is the individual’s visitation frequency of location $i$, $r_{cm}$ is the centre of mass of all visited locations and $N$ is the total number of visited locations. The authors also defined the $k$-radius of gyration, denoted ${r_g}^{(k)}$, which is similar to the overall mobility formula, with the difference that the set of locations $L$ is reduced to the $k$ most visited locations. The value of ${r_g}^{(k)}$ represents the recurrent mobility of the individual. The correlation between the recurrent and the total mobility values distinguishes between the two mobility patterns, namely returners and explorers. If the recurrent mobility of an individual dominates the total mobility, that is ${r_g}^{(k)} > r_g/2$, the individual is classified as a returner. Otherwise, the individual is an explorer.

The authors of \cite{pappalardo2015returners} found that explorers have more impact on disease spread than returners. Their experiments consisted of 10,000 individuals chosen randomly from a pool of 46,000 individuals. To study the impact of each mobility behaviour on the spreading, different proportions of returners and explorers were used. The extent of disease spread is computed through the global invasion diffusion threshold $R*$. This experimental setup presents three main limitations. First, changing the proportion of the mobility groups alters the topology and the characteristics of the network being studied. Second, the contact links connect geographical areas rather then actual human physical encounters. Third, their study of spreading power was performed on a static network in which if a link existed at any point in time that link is considered present during the entire period of study. These limitations make the experiments theoretical as they study a snapshot of a possibly unrealistic contact network. 

Several other limitations emerge when CDR and GPS datasets are used in the context of disease spread \cite{wesolowski2015quantifying}\cite{vazquez2013using}. Most importantly, these datasets lack accurate localization of the individuals due to the distant positioning of cellular towers and poor satellite signals \cite{tizzoni2014use}. Hence, these datasets do not guarantee the existence of real physical encounters between the individuals \cite{tizzoni2014use}. In addition, individuals who are tracked via GPS may be driving a car and hence are not in physical contact with other individuals \cite{tizzoni2014use}. 

Recent studies of epidemiology showed an increasing interest in dynamic networks that guarantee the existence of real physical human contact when studying disease spread \cite{sun2014efficient}. A well suited source of data to study the spreading dynamics of diseases in dense cities are public transit records \cite{browne2016roles}. Several studies have confirmed the presence of a risk factor between the use of bus transportation services and the spreading of many airborne diseases such as tuberculosis, measles and influenza \cite{feske2011giving}\cite{browne2016roles}. The authors of \cite{feske2011giving} stated that bus routes ``are veins connecting even the most diverse of populations'' and showed that individuals who reported regular use of buses are more likely to be infected by tuberculosis. In fact, the congregated and enclosed setting of buses presents a suitable environment for any contagious respiratory disease to spread widely. The infectious pathogens can easily be transmitted onward among passengers through coughing and sneezing \cite{browne2016roles}. In addition, natural and artificial air flow can move suspended pathogens through space. This makes all individuals in an enclosed space like a bus susceptible.

In our previous work \cite{el2019mobility} we confirmed the existence of explorers and returners in the public bus transit dataset of Sydney, Australia. Furthermore, through extensive simulations, we showed that explorers are generally more influential in spreading a disease through the network in comparison to returners. Also, long distance travellers were found to be more influential than short distance travellers. However, when only long distance travellers are considered returners showed a greater propensity in spreading the disease over explorers. The work proved the presence of a deeper and more complex interplay between various mobility aspects when it comes to spreading a disease on a public transport system. In our previous work we did not consider the connectivity aspect of the individuals, which holds critical information in contact based spreading scenarios. Further, our simulations only considered direct encounters between passengers and assumed an infection probability of 1. While disease transmission is possible through direct encounters (i.e. the infected and the susceptible individual are present in the same place at the same time), pathogens can remain in the environment for an extended period of time \cite{shahzamal2018impact}\cite{shahzamal2019indirect}. Therefore, an infectious person can infect susceptible individuals without a direct encounter. Contact networks where only direct encounters are considered are commonly called SPST (same place same time) networks. Networks that in addition to direct contacts capture indirect encounters caused by suspended pathogen are called SPDT (same place different time) networks. Previous studies have shown that considering the suspension time of pathogens changes the underlying topology of the contact network and alters the spreading dynamics of contagious diseases significantly \cite{shahzamal2018impact}\cite{shahzamal2019indirect}.

This paper addresses the limitations of our previous work through the addition of the connectivity dimension and by considering different suspension times of particles and infection probabilities. We identify groups of passengers that have a high potential to spread a disease through a public bus network. Although there are several studies that recognized the importance of human mobility data to identify the most influential individuals in a network, none of the studies tried to use a comprehensive mobility dataset to extract patterns along different movement aspects simultaneously and study the detailed interaction between the different patterns. In particular, we consider the passengers' total mobility, recurrent mobility and connectivity. The impact of each group on the spreading will be evaluated as all the infectious activities occurring in the background of the simulations are traced.

\section{Disease spread modelling} \label{materials and methods}

\begin{figure*}[t!]
\centering
\includegraphics[width=0.9\linewidth]{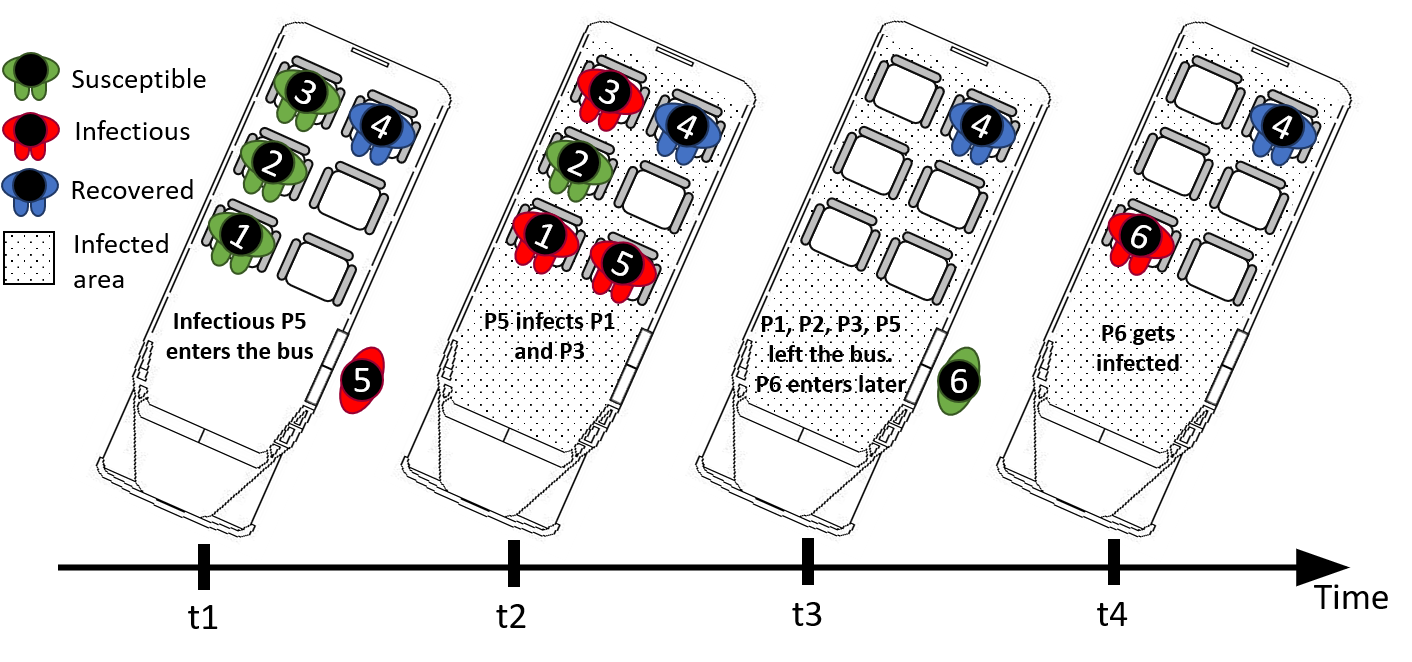}
\caption{The model overview: At time t1, infectious person P5 enters a bus that previously has three susceptible and one recovered individuals embarked. At t2, P1 and P3 get infected through the infectious pathogens disposed by P5 (SPST infection), P2 remains susceptible. At t3, all travellers except P4 have left the bus, however the pathogens disposed by P5 remain inside the bus. P6 enters the bus later and gets infected (SPDT infection). Note that P4 is in the recovered state and has no role in the spreading.}
\label{fig:overview}
\end{figure*}

To understand the disease spread dynamics on a public transport network we construct SPST and SPDT contact networks from the smart card data and run a Susceptible-Infected-Recovered (S-I-R) disease spread model on top. At the beginning of the simulation all bus passengers, except a given number of randomly chosen seed nodes, are susceptible. The seed nodes are infectious and able to transmit the disease to susceptible individuals. When a susceptible individual encounters an infectious individual or, in the case of SPDT networks, comes in contact with pathogens that remain in the environment, the susceptible individual moves to the infectious state with a given probability. The individual remains infectious for a given period of time before recovering from the disease. Once in the recovered state the individual is no longer susceptible and remains in the recovered state until the end of the simulation. Figure \ref{fig:overview} exemplifies the S-I-R disease spread simulation on the bus network.

We demonstrate how the spreading dynamics are affected by changing two key parameters, namely the probability of infection, denoted $\beta$ and the suspension time of pathogens, denoted $d_t$. The case $d_t = 0$ corresponds to an SPST disease spread scenario and hence a susceptible passenger will be infected only if both individuals meet on the same bus at the same time. When $d_t > 0$ the infectious particles remain on the bus for an additional time $d_t$, allowing the infectious passenger to infect susceptible individuals after disembarking.


While simulating the empirical movements of individuals, we track all encounters and infection transmissions. At every encounter the identification number of the two passengers in contact are recorded. Similarly, when an infection is transmitted the identification numbers of the infectious and the susceptible individuals are recorded. To understand how different mobility behaviours influence the transmission paths of the simulated disease we classify the bus passengers into different mobility groups.

We modified the Opportunistic Network Environment (ONE) simulator \cite{keranen2009one} to carry out our trace driven simulations and spread a disease on a large scale real-world transport network.

\subsection{Classification of bus passengers}
Before running the disease simulations on the constructed networks, we cluster the bus passengers into different groups, based on their mobility behaviour. To do so, we simultaneously consider the degree of exploration, the distance travelled and the number of encounters during the period of study. First, we plot passengers' mobility profiles in three-dimensional space, with the $x$-axis corresponding to the passenger's total radius of gyration, the $y$-axis corresponding to the $k$-radius of gyration (i.e. the recurrent mobility) and the $z$-axis corresponding to the number of encounters.

Next, we cluster the individuals into two groups along each dimension. The degree of exploration is divided into returners and explorers, the distance travelled into short distances and long distances and the number of encounters into low connected and highly connected individuals. Classifying our passengers along the three dimensions results in $2^3=8$ different types of movement behaviours. In order to identify each of the groups, we normalize the values of the three dimensions between [0,1] and use the approaches detailed in the following subsections.

\subsubsection{Returners and explorers}
To split the population based on the degree of exploration, we project all the points onto the $xy$-plane and use the bisector method to differentiate between returners and explorers. When plotting the passengers' total mobility and recurrent mobility values on the Cartesian plane, points along the $x$-axis correspond to explorers as their recurrent mobility does not dominate their total mobility and points along the $y = x$ line correspond to returners whose total mobility can be well represented by their recurrent mobility as ${r_g}^{(k)} \approx r_g$. Our clustering approach results in 35.4\% explorers and 64.7\% returners.

\subsubsection{Short distances and long distances}
To cluster the passengers based on their travelled distance, we project the points onto the $x$-axis and apply a standard $K$-means clustering algorithm \cite{pedregosa2011scikit} with $K$=2. This results in two groups, namely, passengers who travel short distances and have a relatively low radius of gyration (87\%) and passengers who travel long distances and have a relatively high radius of gyration (13\%). 

\subsubsection{Low connected and highly connected individuals}
In order to cluster the passengers based on their degree centrality (i.e. the number of encounters) we use a similar approach as in the previous section. We project the points onto the $z$-axis and apply the standard $K$-means clustering algorithm \cite{pedregosa2011scikit} with $K$=2, which splits the population in low connected passengers (39.3\%) and highly connected passengers (60.7\%). Specifically, we differentiate between passengers who encounter a high number of other passengers and those who experience fewer encounters with other passengers during the month of study.

\subsection{Public Transit Traveller Data}
The public transport dataset consists of 20,295,908 trips made by 2 million bus users. The dataset is recorded in the greater Sydney area of New South Wales, Australia during the month of April in 2017. Each trip record records the following information: the passenger's smart card identification, the bus number in use and the time and location the passenger entered and exited the bus. 

\begin{figure}[htbp]
\centering
\includegraphics[trim={0 0 0 0.5cm},clip,width=\linewidth, height=7.5cm]{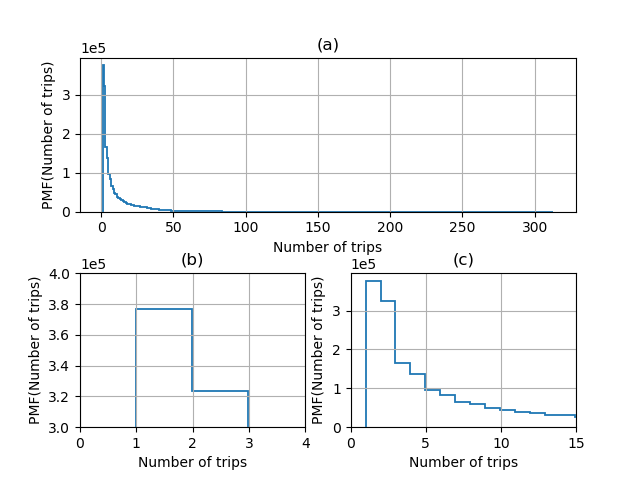}
\caption{(a) The full distribution of trip frequencies. (b) The frequencies for the range [1,2]. (c) The frequencies for the range [1,15].}
\label{fig:dist_trip_freq}
\end{figure}

\begin{figure}
\centering
\includegraphics[trim={0 0 0 0.5cm},clip,width=\linewidth]{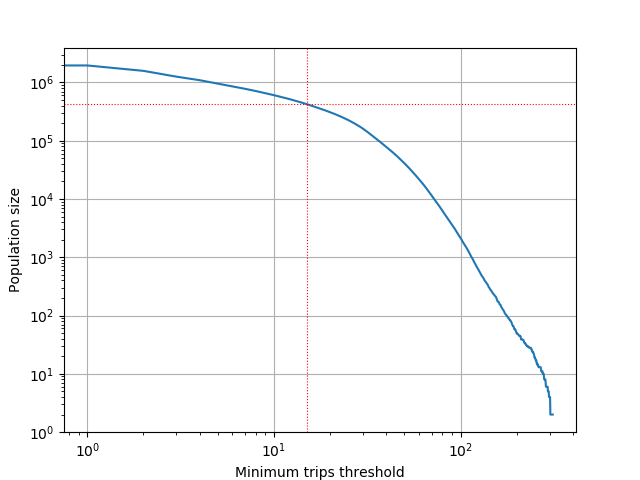}
\caption{The change in the population size with the change of the minimum number of trips made. The dotted red line corresponds to the population size for 15 minimum trips.}
\label{fig:pop_change_with_change_in_threshold}
\end{figure}

Figure \ref{fig:dist_trip_freq} shows the distribution of the trip frequencies for April 2017 indicating that a high number of travellers use Sydney's bus network infrequently during the month of April and few travellers commute by bus on a regular basis. Approximately 700,000 individuals did not travel by bus more than twice during April. These individuals are likely infrequent travellers who use the bus occasionally, visitors who stayed in Sydney for a short time or travellers who lost or damaged their card. As infrequent travellers cannot be classified accurately due to the lack of sufficient data records, we remove these passengers from our analysis. In order to explore how a threshold on the number of trips affects the total number of passengers included in our analysis, we plot in Fig. \ref{fig:pop_change_with_change_in_threshold} the population size against varying threshold values between 1 and the maximum number of trips observed in the data. The population size drops rapidly with the increase of the threshold especially at low values. This is due to the high number of passengers who use the bus only occasionally (see Fig. \ref{fig:dist_trip_freq}). For our analysis, we set the threshold to 15 trips per month. That is, individuals who travelled less than 15 times with the bus during April are excluded from the analysis. The threshold of 15 trips is chosen so that the passengers have travelled at least on half of the days of the month. The final dataset has 36,013,436 records belonging to 424,290 bus passengers. 

\begin{figure}
\centering
\subfigure[]{
    \includegraphics[width=4cm,trim={0 0 2.2cm 0},clip]{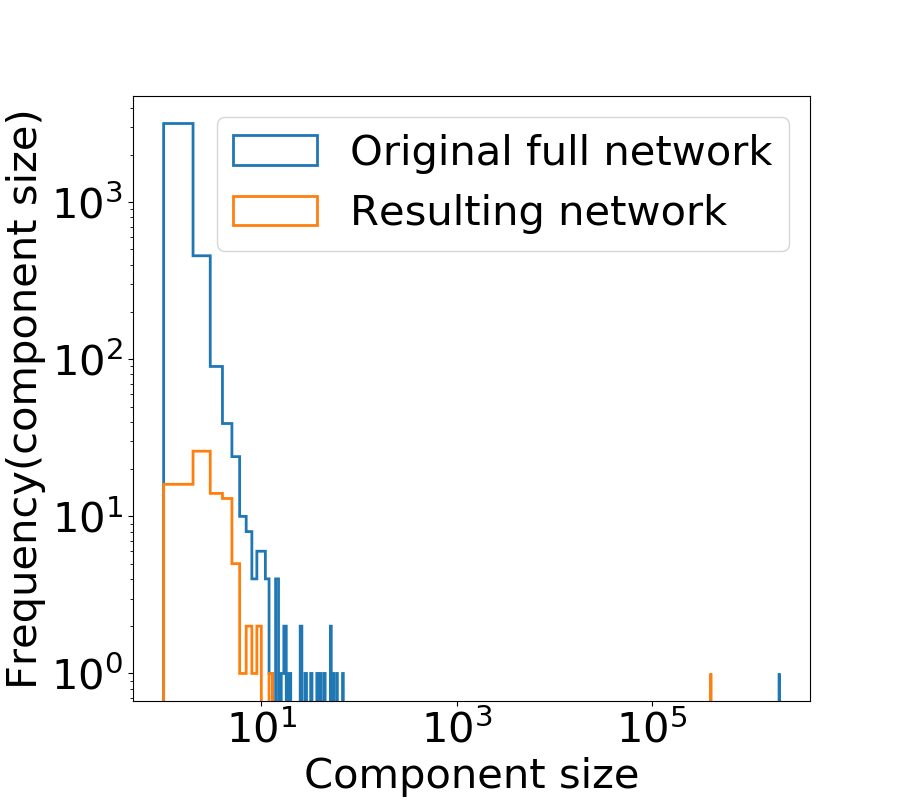}
    
}
\subfigure[]{
    \includegraphics[width=4cm,trim={0 0 2.2cm 0},clip]{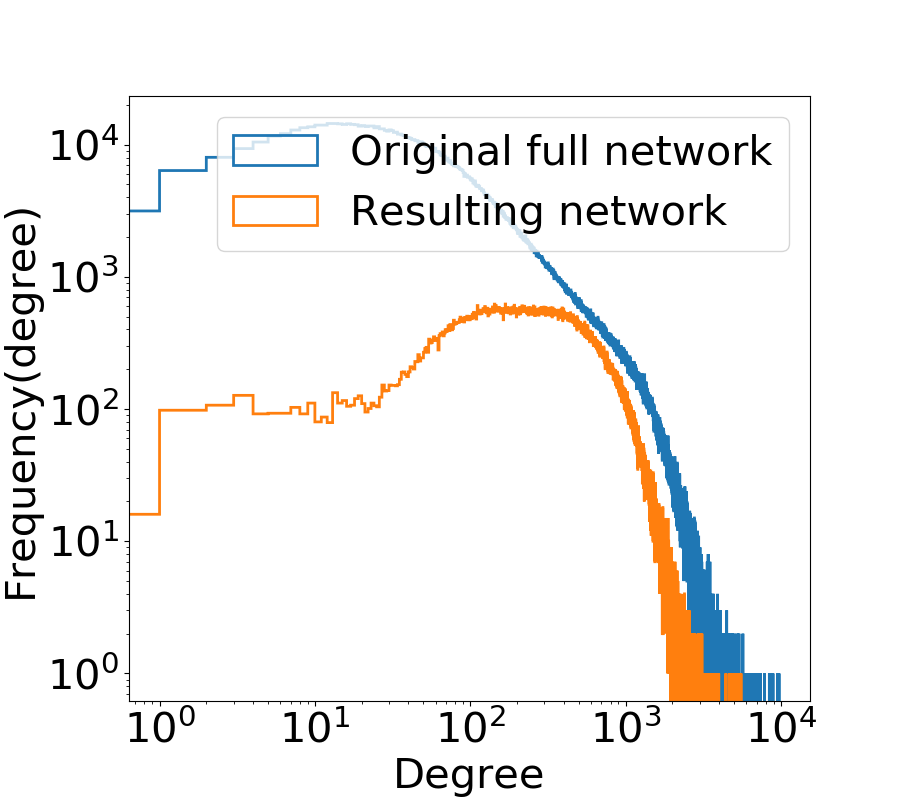}

}
\caption{(a) The frequency distribution of the sizes of connected components. (b) The degree distribution.}
\label{fig:connected_component_analysis}
\end{figure}

\begin{figure*}[ht!]
\centering
\subfigure[The view from the $x$-axis side]{
    \includegraphics[trim={0 0 0 0},clip,width=8cm]{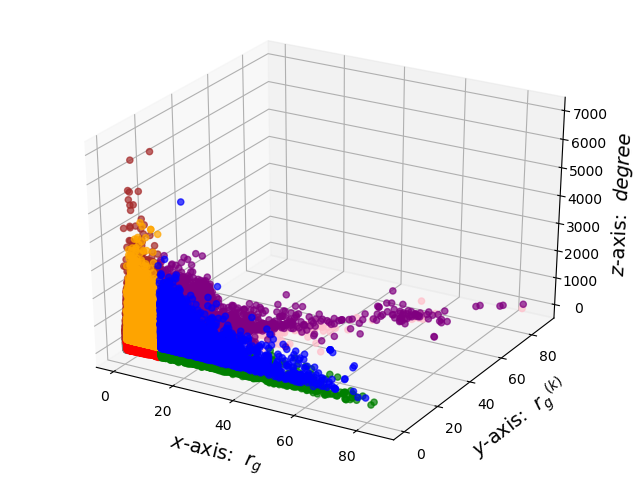}
    
}
\subfigure[The view from the diagonal]{
    \includegraphics[width=8cm]{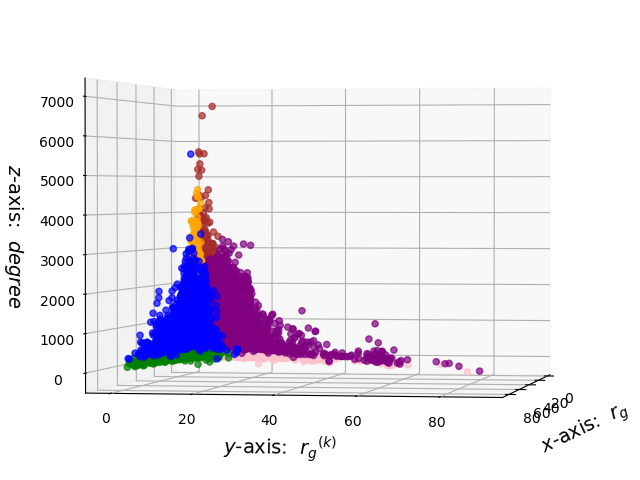}

}

\subfigure[The view from the $y$-axis side]{
    \includegraphics[trim={0 0 0 1.5cm},clip,width=8cm]{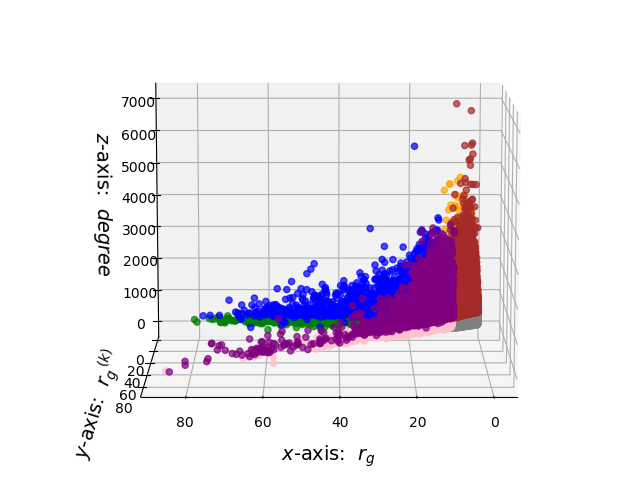}
 
}
\subfigure[top view]{
    \includegraphics[trim={0 0 0 1.5cm},clip,width=8cm]{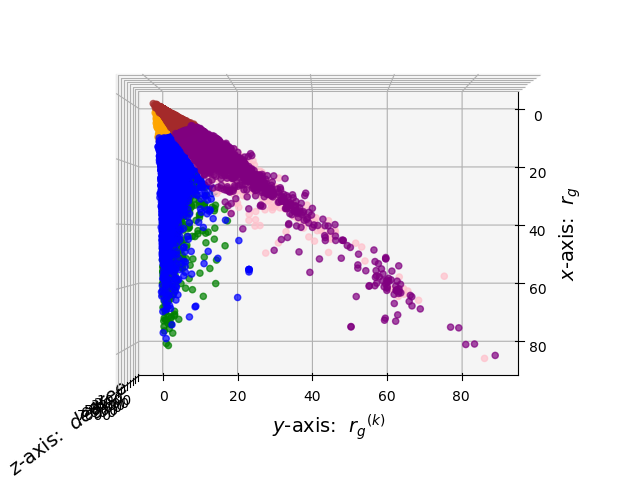}
  
}

\subfigure{
    \includegraphics[width=12cm]{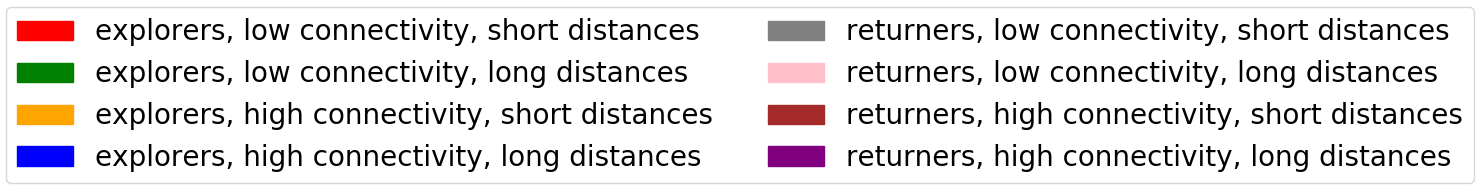}
    
}

\caption{Three-dimensional visualization of the clustered groups.}
\label{fig:3d_representation_full_angles}
\end{figure*}

In Fig. \ref{fig:connected_component_analysis} we compare some key topological aspects of the original and the resulting networks. Figure \ref{fig:connected_component_analysis}.a shows the distribution of connected component sizes for both networks. We notice that the two networks have similar structures, with a giant component and several smaller components. There are fewer components of size one in the resulting network, which can be attributed to the fact that the original network contained many infrequent travellers who were isolated from the rest of the network. This observation also applies to the isolated components consisting of less than eleven individuals.


Figure \ref{fig:connected_component_analysis}.b shows the degree distribution of the original and the resulting networks. The degree of a passenger is the total number of direct encounters experienced during the month of April. We notice that the degree distributions of the original and the resulting networks increase linearly until reaching maximum values of 9,726 and 5,521 respectively. Then both distributions drop exponentially. The drop in the frequencies of the resulting network is due to the removal of passengers who travelled less than 15 trips. The difference between the two distributions is especially clear at low degree values as passengers with few number of trips are less likely to have higher number of contacts.

\section{Results and Discussion} \label{results and discussion}
In this section, we present the identified mobility patterns and discuss the results of our disease spread simulations.

\subsection{Classification results}
The different groups resulting from our classification and clustering tasks are visualized in Fig. \ref{fig:3d_representation_full_angles}. All subfigures display the same plot from a different angle. Each point in Fig. \ref{fig:3d_representation_full_angles} corresponds to one individual in the dataset and its coordinates represent the values of the three-dimensional movement behaviour of the corresponding passenger. In the coming figures and tables we refer to the groups using the notation \{degree of exploration\}\_\{connectivity\}\_\{distance travelled\}.  

\begin{table*}[h]
\small
\centering
\begin{tabular}{ | M{2cm} | M{1.5cm} | M{1.5cm}| M{1.5cm} | M{2cm} | M{2cm} | M{2cm} | } 
\hline
Group & Total number of encounters & Total transmitted infections & Total received infections & Average encounter per individual & Average transmission per individual & Average reception per individual\\ 
\hline
\rule{0pt}{9pt} exp\_high\_long & 4872186 & 17332 & 9625 & 478.415 & 1.701 & 0.945\\ 
\hline
\rule{0pt}{9pt}exp\_high\_short & 33944532 & 90029 & 68844 & 489.685 & 1.298 & 0.993\\
\hline
\rule{0pt}{9pt}exp\_low\_long & 1601377 & 9459 & 12892 & 122.457 & 0.723 & 0.985\\
\hline
\rule{0pt}{9pt}exp\_low\_short & 10126446 & 50881 & 76733 & 130.465 & 0.655 & 0.988\\
\hline
\rule{0pt}{9pt}ret\_high\_long & 11288809 & 27994 & 20946 & 535.801 & 1.328 & 0.994\\
\hline
\rule{0pt}{9pt}ret\_high\_short & 80726848 & 197235 & 158711 & 506.248 & 1.236 & 0.995\\
\hline
\rule{0pt}{9pt}ret\_low\_long & 1012537 & 6544 & 9655 & 129.912 & 0.839 & 0.982\\
\hline
\rule{0pt}{9pt}ret\_low\_short & 16136071 & 94149 & 138217 & 114.791 & 0.669 & 0.983\\
\hline
\end{tabular}
\newline\newline
\caption{Summary of the simulation results showing total and average values of the number of encounters, infections transmitted and infections received for each group at the end of the simulation period.}
\label{tab:my_label}
\end{table*}

\begin{figure}[htbp]
\centering
\includegraphics[trim={0 1cm 0 1cm},clip,width=\linewidth]{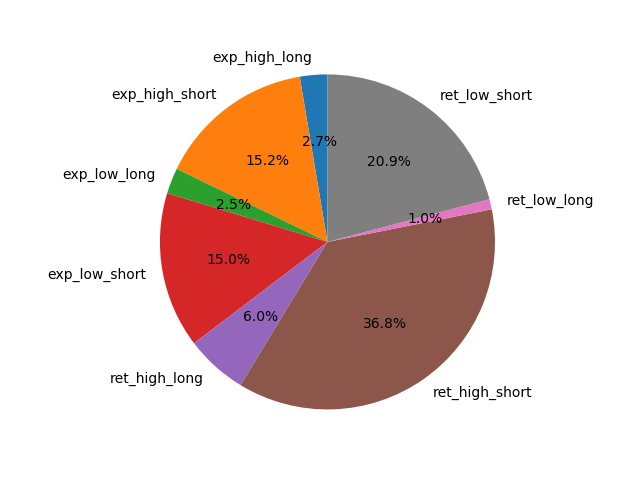}
\caption{A pie chart showing the percentage of each of the eight classified groups of passengers in the network.}
\label{fig:piechart_perc_dist}
\end{figure}

The pie chart in Fig. \ref{fig:piechart_perc_dist} summarizes the percentage of each of the eight classified groups of passengers in the network. We notice that highly connected returners who travel short distances constitute the major portion (36.8\%) of the population. This group of individuals are regular commuters who tend to use public transportation to commute between home and work during peak hours and rarely explore or visit other places during the month. 20.9\% of public transport users are classified as low connected returners who travel short distances. We believe that these individuals regularly travel to specific locations that are less crowded or during off-peak hours, for example people who go to shopping malls in the afternoon. On the other hand, explorers are individuals who in addition to their regular commute visit other places, for example going to malls to shop or going to touristic attractions for leisure.

The following subsections present and discuss our simulation results. For every experiment, we randomly choose 500 individuals who are infectious at the beginning of the simulation and can transmit the disease to susceptible individuals. Individuals remain infectious for five days (which is the average infectious period for influenza \cite{influenzaInfPeriod}) before recovering. Every experiment is simulated 100 times and the results averaged.

\subsection{Experiment 1: SPST scenario} \label{SPST scenario}
In our first experiment we set the infection probability $\beta$ to 1 and the pathogen suspension time $d_t$ to 0, corresponding to an SPST disease spread scenario. This experiment serves as a baseline for comparison to other parameter settings that are explored in further simulations.


In Table \ref{tab:my_label}, we present the total number of encounters and the total number of infections that were transmitted and received by every mobility group. In addition, we compute the average number of encounters, the average number of transmissions and the average number of infections received per individual for each group. Dividing the total number of infections caused by a given group by its population size results in the average number of infections that one individual from that group causes during the simulation period. This average value of infections is not constrained to a specific target group but to all groups in the network. Similarly, dividing the total number of infections received by a given group by its population size, results in the average number of infections that one individual of that group receives during the simulation period. Interestingly, the averages of received infections per individual is nearly the same across all the groups with a value $\simeq 1$. This indicates that at the end of the simulation almost every individual is infected. On the other hand, the average number of infections caused per individual vary from one group to another. Highly connected explorers who travel long distances have the greatest spreading power with an average value of 1.7 transmissions per individual, while low connected explorers who travel short distances have the least spreading power with an average value of 0.6 transmissions per individual. Although highly connected explorers who travel long distances have the highest average infection transmission per individual, this group ranks $4^{th}$ in the average number of encounters per individual. This result shows that it is important to not just divide individuals into explorers and returners, but to distinguish them further along other dimensions such as the distance travelled and the connectivity as their spreading abilities differ.

\begin{figure*}[t!]
\centering
\includegraphics[trim={0 0 0.15cm 0},clip,height=12cm]{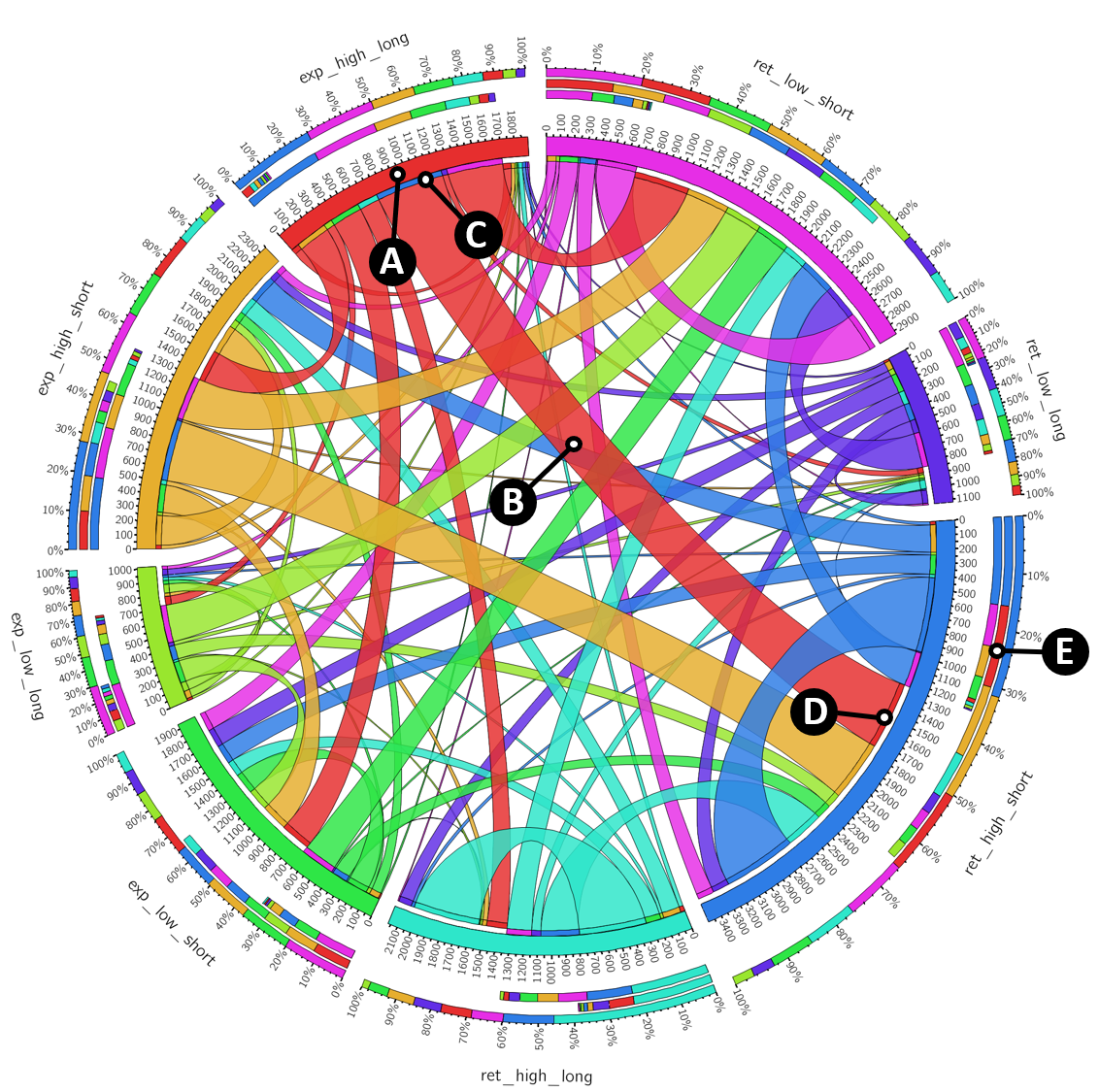}
\caption{A chord diagram showing the disease flows between the different mobility groups resulting from the simulations with parameters $\beta = 1$ and $d_t = 0$. The mobility groups are represented by differently coloured circle segments (e.g. the red segment, labelled ``A'', corresponds highly connected explorers travelling long distances). Links represent disease flow between mobility groups and are coloured by the source group. The link with label ``B'' shows the average number of infections that highly connected explorers who travel long distances transmit to highly connected returners who travel short distances. ``C'' refers to the start of the link coloured by the receiving group. ``D'' refers to the end of the link coloured by the transmitting group. ``E'' refers to the relative infection transmissions, receptions and overall total for each segment.}
\label{fig:chord1}
\end{figure*}

In order to visualise the disease transmission dynamics between the groups we use a chord diagram (see Fig. \ref{fig:chord1}). The diagram shows cumulative disease flows between the different groups. The eight different groups are represented by circle segments, with each group being associated to a unique colour. For example, the red segment corresponds to highly connected explorers who travel long distances (see label ``A'' in Fig. \ref{fig:chord1}). The links indicate the volume of disease transmissions between any two groups and are assigned the same colour as the source group. The thickness of each link is proportional to the average number of people that one individual from the source group infects in the target group. For example, the link labeled ``B'' shows the volume of disease flow transmitted from the group of highly connected explorers who travel long distances (red segment) to the group of highly connected returners who travel short distances (blue segment). Links that start and end at the same segment represent disease transmissions between individuals of the same mobility group. For scaling purposes, we multiply all average number of infections caused per individual by 1000 and show the resulting values in the chord diagram.



The diagram in Fig. \ref{fig:chord1} clearly identifies four dominant infection paths. These occur amongst highly connected returners who travel long distances (cyan segment) with individuals causing on average 0.52 infections within their own mobility group and highly connected returners who travel short distances (blue segment) with individuals causing on average 0.57 infections within their own mobility group. Highly connected explorers who travel long distances (red segment) infect on average 0.5 highly connected returners who travel short distances (blue segment) during the simulation period. Highly connected explorers  who travel short distances (orange segment) infect on average 0.47 highly connected returners who travel short distances (blue segment).
Furthermore, low connected returners who travel short distances form a group that is prone to receive infections, but less likely to infect individuals from other groups (see the incoming non-pink links that occupy the majority (77\%) of the pink segment in Fig. \ref{fig:chord1} and Table \ref{tab:my_label}). 
On the other hand, highly connected explorers who travel long distances have caused the greatest number of infections per individual on average. However, this group is less likely to get infected in comparison to other groups (see red segment in Fig. \ref{fig:chord1}). We observe that the disease transmissions from highly connected explorers who travel long distances dominate this group's activity as the red outgoing links going to all other groups constitute the majority of the segment with more than 90\%. That is, even a low number of infected individuals of this group would be sufficient to infect other groups and spread the disease through the entire network. Highly connected explorers who travel long distances infect 1.7 individuals on average during the simulation period. 
Highly connected returners who travel short distances receive a high number of infections and mostly infect individuals within their own group (see blue segment in Fig. \ref{fig:chord1}). This behaviour is expected, as this group consists of regular commuters who display consisted movement behaviour. Highly connected returners who travel long distances (see cyan segment in Fig. \ref{fig:chord1}) display a similar behaviour of mostly infecting individuals within their own group. Highly connected explorers who travel long distances spread infections to all other groups (see red segment in Fig. \ref{fig:chord1}), although the average number of encounters is lower than groups who infect specific target groups.

Our results highlight important interactions between the eight identified groups and shed light on disease spread dynamics that should be given more attention while monitoring a disease and applying prevention measures. To understand how different disease types and characteristics change the spreading dynamics between the eight groups, we perform two additional experiments.

\subsection{Experiment 2: SPDT scenario}
In this experiment, we run the simulations with different suspension times of pathogens, i.e. $d_t =$ 15, 30, 60 and 120 minutes, while keeping $\beta = 1$. For each value of $d_t$ we construct a matrix that shows the difference in the average number of infections caused and received by each mobility group in comparison to experiment 1 ($\beta = 1$, $d_t = 0$). Positive values refer to a gain in disease transmissions, whereas negative values indicate a loss. The rows of the matrix correspond to the groups that cause the infections and the columns correspond to the groups that receive the infections.

\begin{figure}
\centering
\subfigure[]{
    \includegraphics[width=\linewidth]{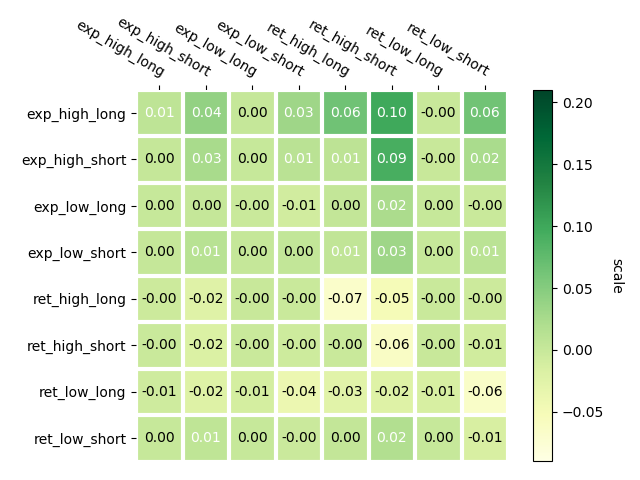}
    
}
\subfigure[]{
    \includegraphics[width=\linewidth]{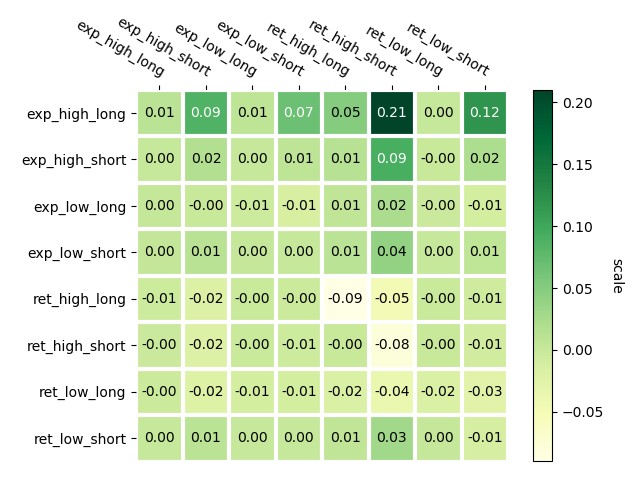}

}

\caption{Matrix representations showing the difference in the average number of infections caused when the suspension time is changed: (a) from 0 to 15 minutes, (b) from 0 to 30 minutes.}
\label{fig:delay_changes_subfigs}
\end{figure}

Figures \ref{fig:delay_changes_subfigs}.a and \ref{fig:delay_changes_subfigs}.b show the matrices for a suspension time of 15 minutes and 30 minutes, respectively. The matrix in Fig. \ref{fig:delay_changes_subfigs}.a shows that the average number of infections caused per individual increased or remained the same for the four groups of explorers (top four rows of the matrix). The spreading potential of the four groups of returners (bottom four rows of the matrix) generally decreased. As the suspension time is increased to 30 minutes, we observe further increases in the average number of infections caused by explorers and further decreases for returners (see Fig. \ref{fig:delay_changes_subfigs}.b). We highlight that the increase in $d_t$ weakens the spread of infections within (self-loops) the two groups of highly connected returners. The loss in the infection power of returners coincides with an increase in infections caused by explorers, especially for highly connected explorers who travel long distances. Since almost every individual of the population is infected at the end of the simulation period, we conclude that an increase in the time that pathogens remain in the environment favours the infection power of explorers. That is, explorers are even more influential in an SPDT disease spread scenario. We only show the results for $d_t =$ 15 and 30 as no change in the behaviour was seen for $d_t =$ 60 and 120, the values for explorers keep increasing and those of returners decrease.

\subsection{Experiment 3: impact of the infection probability}

In the third experiment, we vary the infection probability $\beta$, while setting the pathogen suspension time $d_t = 0$. The considered probabilities are 0.05, 0.1, 0.15, 0.25, 0.5, 0.75 and 1. 

\begin{figure}[htbp]
\centering
\includegraphics[width=\linewidth]{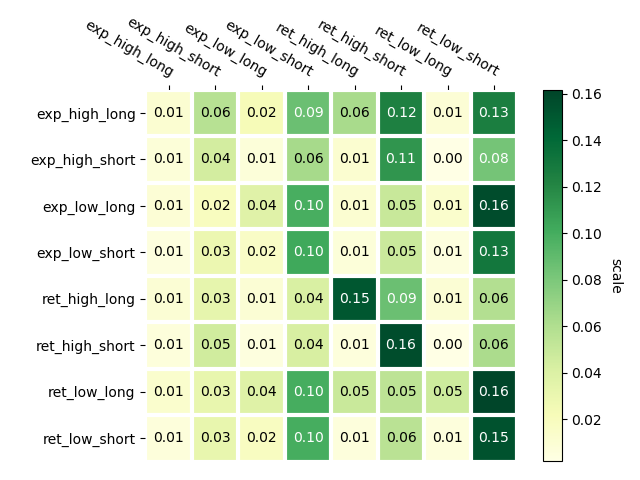}
\caption{A matrix representation showing the difference in the average number of infections caused when the infection probability is changed from 0.05 to 0.25.}
\label{fig:infectionProb_btw_0.05_0.25}
\end{figure}

To understand the effect of the infection probability on the disease spread we construct matrices that show the differences in the average number of infections caused per individual between each two consecutive values of $\beta$. The average number of infections caused per individual increases rapidly for all groups with an increase of $\beta$ from 0.05 to 0.25. This result is visualised in Fig. \ref{fig:infectionProb_btw_0.05_0.25} with all matrix elements being positive. When $\beta$ is increased from 0.25 to 0.5 we see only a slight increase in the spreading power of all groups. Further increases of $\beta$ to 0.75 and 1 do not result in significant changes in spreading powers. 


Figure \ref{fig:infectionProb_btw_0.05_0.25} shows that all individuals who travel short distances experienced the most increase in the number of received infections, whether they are low or highly connected. This pattern can be seen through the dark coloured columns of the short distances groups. The observation is due to these groups constituting the highest percentages in the network allowing them to have the highest total number of encounters (see Fig. \ref{fig:piechart_perc_dist} and Table \ref{tab:my_label}). In addition, the increase in the probability of infection strengthens the self-loops of the groups (infections within the same group), especially those of the short distance returners. We conclude that increasing the infection probability favours the spreading power all mobility groups. The increase of spreading power is relative to the interaction between each pair of mobility groups. For increasing infection probabilities each element in the matrix increases until ultimately reaching the values presented in Section \ref{SPST scenario} in which the probability is set to 1.

\section{Conclusion} \label{conclusion and future work section}
This is the first study to identify mobility patterns along three dimensions simultaneously, namely the degree of exploration, the distance travelled and the number of encounters. We found previously unknown mobility patterns that were thoroughly investigated to understand the spreading dynamics of contagious diseases on a city wide public transport system. We ran extensive disease spread simulations with varying values for the infection probability and the suspension time of pathogens. Our results show that characterizing individuals along multiple dimensions simultaneously uncovers a complex infection interplay between the different groups of travellers. Furthermore, the infection probability and the suspension time of pathogens play different roles in the spread. Highly connected passengers who regularly return to the same places play the most important role in the spreading when pathogens do not remain in the environment. However, with an increase in the suspension time of pathogens, highly connected passengers who visit new locations are the most influential. Unlike the suspension time, increasing the infection probability does not affect particular mobility groups, but increases the infection power of all groups especially for returners who travel short distances.
Our simulation experiments are abstractions of the real-world and flexible to adapt to different contexts. We presented a framework that can be applied to model any disease that is spread through a physical contact network. Our findings are especially beneficial to advise health authorities on the design of more efficient intervention and containment strategies depending on the characteristics of the emerging diseases. We plan to open-source the modified simulator in order to be used broadly for similar types of datasets and scenarios.

\medskip

\bibliographystyle{unsrt}
\bibliography{mybibliography}

\end{document}